# Risk Assessment Techniques and Survey Method for COTS Components

Rashmi Gupta[1] and Shalini Raghav[2]

[1]Department of Computer Science, RGPV University, TIT, Bhopal

`rgupta6773gmail.com`

[2]Department of Computer Application, GBTU University, HRIT, Ghaziabad

`shalini1182@gmail.com`

## Abstract

*The Rational Unified Process a software engineering process is gaining popularity nowadays. RUP delivers best software practices for component software Development lifecycle It supports component-based software development. Risk is involved in every component development phase .neglecting those risks sometimes hampers the software growth and leads to negative outcome. In Order to provide appropriate security and protection levels, identifying various risks is very vital. Therefore Risk identification plays a very crucial role in the component based software development This report addresses incorporation of component based software development cycle into RUP phases ,assess several category of risk encountered in the component based software. It also entails a survey method to identify the risk factor and evaluating the overall severity of the component software development in terms of the risk. Formula for determining risk prevention cost and finding the risk probability is also been included. The overall goal of the paper is to provide a theoretical foundation that facilitates a good understanding of risk in relation to component-based system development.*

## 1 .Introduction

Component-based software development (CBSD) is an emerging development paradigm that promises to accelerate software development and reduces development costs by assembling systems from pre-fabricated components [2,14] .Component based development provides the way of purchasing the components from the market rather than building the components from scratch [14]. It facilitates two techniques "Either building new components in house or buying the components from the third parties". By using Component-based software development (CBSD) development time of the software decreases dramatically, leads to increase in the usability of the products, and decrease in the production costs [13] .CBSE liberated  the programmer from thinking about details, as it shifts the emphasis from programming to composing software systems using several components [17].Component-based software development provides a rapid mechanism for increasing the functionality and efficiency of a system, But component-based development carries significant risk throughout the system life cycle. Rational Unified Process is a software Engineering process that supports component based development activities. Component-based software development faces several risks during the entire software development life cycle. These risks are associated with the behavior of COTS, vendor support, technologies and the development process [15]. Risk is a factor that involves uncertain danger and can obstruct the development of the software. Therefore risk identification techniques and categorizing risk across the several phases of the component based software development are extreme important so that the severity of the risk can be reduced. Risk identification mechanism plays a vital role in estimating the probability of the risk occurrence

## 1.1 Components

**Software component** is a module that encapsulates related data and functions, software package or can be the web service. It can also be used as a building block to create larger, more complex software systems. Data and various functions encapsulated in the components are semantic related to each other. Components interact as well as use the services of each other through interfaces. Inner functionality or the structure of the components is encapsulated or is not known to the client [10].

## 1.2. RUP Phases

**1.2.1. Inception phase** is the first phase of the RUP that defines objectives as well as the scope of the project. In this phase the features like planning project, risk assessment techniques and project description features like requirements of the project, various checks are established. It deals with recognition of the requirements of the users [11]. Various quality levels are defined establishment of the cost and budget is also done during this phase. During this phase a baseline will be established that will compare actual expenditures versus planned expenditures .Business plan like market research, business context is also defined during this phase A basic use case model that define functionality of the system is generated .Inception phase incorporate component selection activity process [2] that encompass market survey for finding out the appropriate components from the vendor.

**1.2.2. Elaboration Phase** is the second phase of the RUP that defines architecture of the software. The elaboration phase is where the project starts to take shape.[11] In this phase the problem domain analysis is made. Various use cases diagrams along with the use cases and the actors are identified. Generation of the development plan for the overall project occurs. Prototypes are generated. Components interact with each other through the interfaces. Creating Well defined architecture of the interfaces is done in this phases. Appropriate process model for component development will be taken into consideration

**1.2.3 .Construction Phase** is the third phase of the RUP that produces the first external release of the software [11]. This phase encompass component integration activity. Component integration [2] is possible through coding. Maximum coding is done in this phase.

**1.2.4. Transition phase** is the last phase of the RUP in which system is made available to the end users [11]. Various training programs regarding the use of the system and about the used technologies are conducted .beta testing is conducted at the user site to validate the system. Validation of the quality levels that were defined during the inception phase is also done. This phase incorporates the component evolution activity [2].

## 2. Related Work

In this section we place our work in relation to ongoing research within related area. Component model has been proposed that serves as a foundation for component-based software risk analysis by integrating several component risks as part of the component behavior. In component-based software risk analysis risks are identified, analyzed and then documented. Result of risk analysis at component level is compos able. Techniques for component-based risk analysis facilitate the integration of risk analysis into component-based development, and thereby make it easier to predict that how upgrading subparts influences component risks. Various studies on occurrence of typical risks in component based development and comparing the effectiveness of performed risk-reduction activities has been explored. Several efficient risk reduction activities, such as carefully examining the quality of component in selection phase, estimating the behavior of the components, integrating unfamiliar components first, and monitoring the reputation of component vendor has been explored. In addition, several context variables, such as number of components, experience of the developer and quality requirements of the system, have been discovered as confound factors on the relationships between risk reduction activities and the occurrence of certain risk items. Large number of international survey across different countries based on risk identification, risk management in using COTS components is reported upon and discussed. The survey investigated several risk-management activities and their correlations with occurrences of different risks in component-based software development. Results also illustrate several effective risk reduction activities, such as conducting integration testing early and incrementally, evaluating various tools that supports the component development activities, and monitoring the capability of components vendors Lot of research covering the component activity areas associated with the different risks has been conducted. Identification, assessment of risk, sensitivity of risks and the impacts of different risks on the different nature of projects is also been conducted. Several mitigation as well as remedial steps for these risks has been defined. Various empirical studies on how process improvement and risk management activities were performed .The results of the empirical study reveal that allocating more effort into learning OTS components, performing the integration testing early, estimating the behavior and the specifications of the components ,inspecting the quality of components at each phase can reduce the risks up to maximum extent .CBSRAM is developed which is used to categorize component And help in selecting suitable components based on the risk measures. The quantitative framework suggested for CBS analyzes conflicting components Various Papers related to the categorization of the risks in component-based development has been discussed. In conventional risk analysis the portions of the environment that are important for estimating the risk-level is also analyzed .Conventional risk analysis approach consists of: (1) a framework for CBRA (2) a modular Risk modeling; (3) a formal foundation for modular risk modeling. The framework for component-based risk analysis provides a technique for analyzing system component wise and then combining the result at each component level. Modular risk modeling involves 1) decomposing a system into components 2) composing components to form a larger system. Modular risk modeling approach analyzes and model different risks and aid in identifying, analyzing and documenting the several component risks. The component based model provides a foundation for integrating risk analysis into component-based development. Component-based risk analysis provides a framework for conducting risk analysis at the component level. Framework is based on the CORAS method (graphical risk modeling language). Modular risk modeling introduces the risk graph, which is act as an abstraction of several risk modeling techniques, such as tree-based diagrams. A denotation model for component-based risk analysis has been developed that represent the behavior of a component by a probability distribution over communication histories

## 3. Risk Identification during component development activities in RUP Phases

Risk identification in component based software is the technique that is used for identifying the various types of risk at every phase of the development. It is done at the component level and the risk identified at each component level will be added to the component at the next level The objective of component-based risk identification and categorization is done in order to develop reliable and trustworthy components [4].

### 3.1. Component based software development Risk during inception phase

- Prerequisite Quality is not met due to the lack of market survey [9] that has to be done to know the requirement of the customer
- When the COTS Component and the requirement suggested by the user does not matches
- Requirements of the users changes frequently
- Budgets and schedules are not realistic
- Unclear requirements specification
- Lack of accuracy in schedule
- Lack of reliable and suitable licensing [9] contracts that encompass the appropriate documentation and responsibility of the vendor and developer in case of failure
- Rigidness in Time constraints of schedule generates inflexibility
- Market survey is not done properly than appropriate components[9] that can map with user requirements cannot be found
- Lack of contingency planning
- Rapid requirement change of the user
- Component search suffers from appropriate fetching [9] and classification mechanism provided by the marketers
- Architectural prototype not defined properly
- Latest Technologies and fresh arrived COTS product are not analyzed or lack of market survey
- Vendors incapability to delivers mind blowing demos and specifications of the COTS product
- Architecture was not analyzed during the component selection process [8]
- Cumbersome and complicated requirements
- Lack of vendor support
- Missing authenticity of the components due to the lack of certified components
- Unavailability of the source code leads to judging nature and the behavior of the components
- Inappropriate domain knowledge of developer [7].

### 3.2. Component based software Development Risk during elaboration phase

- Higher Complexity of components architecture and the connectors introduces the chances of risk [1].
- Mismatch between connectors and message protocols
- Interface specification of the components is not clear or not specified [5] properly
- Incompatible or mismatch Interfaces may obstruct the data communication between the components which wants to exchange data
- Use of the Software model that does not support component based software development process
- False assumption of the internal structure and internal specification made by the COTS component about each other
- Lack of resilient architecture
- Existence of the loop holes in the architecture review process
- Components are not platform independent
- Lack of executable architectural prototype
- Mismatch occurrence between planned expenses and actual expenses
- Security aspects are not considered and the vulnerability of the components is very high
- Prototypes that demonstrably mitigate each identified technical risk are not defined

- Components are not interoperable[5] with each other due to missing well defined interfaces
- Lack of Software architecture document that is extreme crucial in order to gain knowledge about the component
- Loop holes in Architectural Style [13] as architects have at hand incomplete, imprecise and uncertain component information [16].
- Component architecture are not compatible with each other thus makes integration of the component tough
- Component based software prototypes cannot be realized in early phases of the software cycle make architecture verification of the interfaces difficult

### 3.3. Component based software development Risk during construction phase

- Wrong interface construction may hinder the proper flow of the information or data between several components
- Development of the wrong functions at the time of coding leads to several exceptions
- Lack of regular watch on the component based development process generate several problems
- lack of test suites and test cases that facilitates coordination among the component
- Generation of Incompatibility between user requirements stated earlier in the component based system and the new versions [3] developed.
- Staffs persons indulge in integration process of the components are not technological sound [1].
- Behavior of the components cannot be judged in component based development due to the absence of the availability of the code of the component
- Lack of Technology expertise and poor work knowledge and skills of assembler leads to Poor Component evaluation and integration [1].
- Missing compatibility between the different versions [3] of the component based software.
- Existence of Poor or no documentation feature for the new versions [3].
- Poor stability control -If the stability is not incorporated in the component based system then
- Doing change in one component will make a heavy impact on the other component
- Unavailability of the competent staff
- Unavailability of the internal structure of the component makes the testing process tough and unreliable.

### 3.4. Component Based Software Development Risk during transition phase

- End user training sessions are not conducted
- Component based software that is developed cannot accommodate changes preferred By the user
- Occurrence of incompatibility between the component based product being developed And the quality level that has been set during the initial phase of the software development cycle
- Complicated system manual results lack of understanding by the users
- Quality services after the COTS software installation at the user site are not given
- User is not facilitated with the upgraded copies of the component based software
- Updating or alteration of the component based system cannot be facilitated
- Lack of tracing of alternate component in case of failure [8]
- Planning the maintenance is difficult as the components have asynchronous cycle[8]

## 4. Risk Probability and Risk prevention cost formula

Risk Prevention Cost $(_{Rpc})$= Cost Of Preventing Threats(CPT)+ Cost Of Preventing vulnerabilities(CPV)+ Quality Appraisal Cost(QAC)+(External/Internal)Failure Cost(FC)

Cost Of Preventing Threats (CPT) = $\sum_{i=1}^{n}(Ti * CTi)$

Where Ti=Similar Number of Threat

$C_{Ti}$=cost of the particular Threat

Cost of preventing vulnerabilities (CPV) = $\sum_{i=1}^{n}(Ti * CVi)$

Where $_{Ti}$=Similar Number of vulnerabilities

$C_{Vi}$=cost of the particular vulnerability

Risk Prevention Cost $(_{Rpc})$ = CPT+CPV+QAC+FC

Risk Probability= (CPT+ CPV) ÷ $(_{Rpc})$

Where CPT= Cost Of Preventing Threats

Where CPV=Cost of Preventing vulnerabilities=

Where RPC =Risk Prevention Cost

## 5. Risk Table showing Risk factor associated with each development phase

| Feature | Phases | Risk Encounter | Users | | Risk Manager | | Total | | Average | RVR | Priority |
|---|---|---|---|---|---|---|---|---|---|---|---|
| | | | Total User. | Sum Of Rating | Total Q.M. | Sum Of Rating | No. | Sum | | 31.45 | |
| Risk | Inception | A | 5 | 9 | 6 | 8.5 | 11 | 17.5 | 1.59 | | 7 |
| | | B | 5 | 8 | 6 | 9 | 11 | 17 | 1.54 | | |
| | | C | 5 | 9 | 6 | 11 | 11 | 20 | 1.81 | | |
| | | D | 5 | 9.5 | 6 | 8.5 | 11 | 18 | 1.63 | | |
| | | E | 5 | 6 | 6 | 9 | 11 | 15 | 1.36 | | |
| | | F | 5 | 10 | 6 | 9 | 11 | 19 | 1.68 | | |
| | | G | 5 | 8 | 6 | 8.5 | 11 | 16.5 | 1.72 | | |
| | | H | 5 | 8 | 6 | 11 | 11 | 19 | 1.27 | | |
| | | I | 5 | 7 | 6 | 6 | 11 | 13 | 1.22 | | |
| | | J | 5 | 7 | 6 | 6.5 | 11 | 13.5 | 1.31 | | |
| | | K | 5 | 7 | 6 | 7.5 | 11 | 14.5 | 1.27 | | |
| | | L | 5 | 10 | 6 | 7 | 11 | 17 | 1.45 | | |
| | | M | 5 | 9 | 6 | 6 | 11 | 15 | 1.40 | | |
| | | N | 5 | 8 | 6 | 6.5 | 11 | 14.5 | 1.5 | | |
| | | O | 5 | 10 | 6 | 8.5 | 11 | 18.5 | 1.68 | | |
| | | P | 5 | 10 | 6 | 9 | 11 | 19 | 1.72 | | |
| | | Q | 5 | 7 | 6 | 6 | 11 | 13 | 1.18 | | |
| | | R | 5 | 10 | 6 | 6.5 | 11 | 16.5 | 1.5 | | |
| | | S | 5 | 8.5 | 6 | 6 | 11 | 14.5 | 1.31 | | |
| | | T | 5 | 6.5 | 6 | 11 | 11 | 17.5 | 1.59 | | |
| | | U | 5 | 10 | 6 | 6 | 11 | 16 | 1.45 | | |

| | | | | | | | | | | |
|---|---|---|---|---|---|---|---|---|---|---|
| | Elaboration | V | 5 | 8.5 | 6 | 8.5 | 11 | 17 | 1.54 | |
| | | W | 5 | 5 | 6 | 6.5 | 11 | 11.5 | 1.04 | |
| | | X | 5 | 9.5 | 6 | 6 | 11 | 15.5 | 1.40 | 22.43 | 5 |
| | | Y | 5 | 7 | 6 | 6 | 11 | 13 | 1.18 | |
| | | Z | 5 | 6 | 6 | 11 | 11 | 17 | 1.54 | |
| | | AB | 5 | 8.5 | 6 | 6.5 | 11 | 15 | 1.36 | |
| | | CD | 5 | 6 | 6 | 7 | 11 | 13 | 1.18 | |
| | | EF | 5 | 7 | 6 | 7.5 | 11 | 14.5 | 1.31 | |
| | | GH | 5 | 5.5 | 6 | 7.5 | 11 | 13 | 1.18 | |
| | | IJ | 5 | 7.5 | 6 | 9 | 11 | 16.5 | 1.5 | |
| | | KL | 5 | 6.5 | 6 | 9 | 11 | 15.5 | 1.40 | |
| | | MN | 5 | 5 | 6 | 11 | 11 | 16 | 1.45 | |
| | | OP | 5 | 7.5 | 6 | 8.5 | 11 | 16 | 1.45 | |
| | | QR | 5 | 6.5 | 6 | 6.5 | 11 | 13 | 1.18 | |
| | | UV | 5 | 5 | 6 | 6 | 11 | 11 | 1 | |
| | | WX | 5 | 5 | 6 | 6.5 | 11 | 11.5 | 1.04 | |
| | | YZ | 5 | 9.5 | 6 | 9 | 11 | 18.5 | 1.68 | |
| | Construction | ABC | 5 | 9.5 | 6 | 9 | 11 | 18.5 | 1.68 | |
| | | DEF | 5 | 5 | 6 | 6 | 11 | 11 | 1 | 17.68 | 3 |
| | | GHI | 5 | 5 | 6 | 11 | 11 | 16 | 1.45 | |
| | | JKL | 5 | 6.5 | 6 | 8.5 | 11 | 15 | 1.36 | |
| | | MNO | 5 | 6.5 | 6 | 6 | 11 | 12.5 | 1.13 | |
| | | PQR | 5 | 6 | 6 | 6.5 | 11 | 12.5 | 1.13 | |
| | | STU | 5 | 6 | 6 | 6.5 | 11 | 12.5 | 1.13 | |
| | | VWX | 5 | 6 | 6 | 6 | 11 | 12 | 1.09 | |

| | | | | | | | | | |
|---|---|---|---|---|---|---|---|---|---|
| | YZA | 5 | 6 | 6 | 6 | 11 | 12 | 1.09 | |
| | ABCD | 5 | 7.5 | 6 | 7.5 | 11 | 15 | 1.36 | |
| | EFGH | 5 | 6.5 | 6 | 6 | 11 | 12.5 | 1.13 | |
| | IJKL | 5 | 6.5 | 6 | 7.5 | 11 | 14 | 1.27 | |
| | MNOP | 5 | 6.5 | 6 | 7.5 | 11 | 14 | 1.27 | |
| | QRST | 5 | 8.5 | 6 | 9 | 11 | 17.5 | 1.59 | |
| Transition | AF | 5 | 10 | 6 | 6 | 11 | 16 | 1.45 | |
| | CF | 5 | 10 | 6 | 6.5 | 11 | 16.5 | 1.5 | 11.3 7 |
| | BF | 5 | 5 | 6 | 7.5 | 11 | 12.5 | 1.13 | 2 |
| | EF | 5 | 6 | 6 | 6 | 11 | 12 | 1.09 | |
| | GF | 5 | 6 | 6 | 6.5 | 11 | 12.5 | 1.13 | |
| | KF | 5 | 7 | 6 | 6.5 | 11 | 13.5 | 1.22 | |
| | LF | 5 | 7 | 6 | 8.5 | 11 | 15.5 | 1.40 | |
| | MF | 5 | 5.5 | 6 | 7.5 | 11 | 13 | 1.18 | |
| | TF | 5 | 7.5 | 6 | 6.5 | 11 | 14 | 1.27 | |

Table 1 Survey table For Evaluating Risk Probability

Table 2 Description Of abbreviated symbols for Risk are described as follows

| A | When the COTS Component and the requirement suggested by the user does not matches |
|---|---|
| B | Requirements of the users changes frequently |
| C | Budgets and schedules are not realistic |
| D | Unclear requirements specification |
| E | Lack of accuracy in schedule |
| F | Lack of reliable and suitable licensing contracts that encompass the appropriate documentation and responsibility of the vendor and developer in case of failure |
| G | Rigidness in Time constraints of schedule generates inflexibility |
| H | Market survey is not done properly than appropriate components that can map with user requirements cannot be found |
| I | Lack of contingency planning |
| J | Rapid requirement change of the user |
| K | Component search suffers from appropriate fetching and classification mechanism provided by the marketers |
| L | Architectural prototype not defined properly |
| M | Latest Technologies and fresh arrived COTS product are not analyzed or lack of market survey |
| N | Vendors incapability to delivers mind blowing demos and specifications of the COTS product |
| O | Architecture was not analyzed during the component selection process [8] |
| P | Cumbersome and complicated requirements |
| Q | Lack of vendor support |
| R | Missing authenticity of the components due to the lack of certified components |
| S | Unavailability of the source code leads to judging nature and the behavior of the components |
| T | Inappropriate domain knowledge of developer [7] |
| U | Higher Complexity of components architecture and the connectors introduces the Chances of risk |
| V | Mismatch between connectors and message protocols |
| W | Interface specification of the components is not clear or not specified properly |
| X | Incompatible or mismatch Interfaces may obstruct the data communication between the components which wants to exchange data |
| Y | Use of the Software model that does not support component based software Development process |
| Z | Prerequisite Quality is not met due to the lack of market survey that has to be done to Know the requirement |
| A B | False assumption of the internal structure and internal specification made by the COTS component about each other |
| C D | Lack of resilient architecture |
| E F | Existence of the loop holes in the architecture review process |
| G H | Components are not platform independent |
| IJ | Lack of executable architectural prototype |
| K L | Mismatch occurrence between planned expenses and actual expenses |
| M N | Security aspects are not considered and the vulnerability of the components is very high |
| O P | Prototypes that demonstrably mitigate each identified technical risk are not defined |
| Q R | Components are not interoperable with each other due to missing well defined interfaces |
| U V | Lack of Software architecture document that is extreme crucial in order to gain Knowledge about the component |
| W X | Component architecture are not compatible with each other thus makes integration of the component Tough |
| Y Z | Component based software prototypes cannot be realized in early phases of the software cycle make architecture verification of the interfaces difficult |
| A B C | Wrong interface construction may hinder the proper flow of the information or data between components |
| D E F | Development of the wrong functions at the time of coding leads to several exceptions |
| G HI | Lack of regular watch on the component based development process generate several problems |
| J K L | Lack of test suites and test cases that facilitates coordination among the component |
| M N O | Generation of Incompatibility between user requirements stated earlier in the component and new versions developed |
| P Q R | Staff persons indulge in integration process of the components are not technological sound |
| S T U | Behavior of the components cannot be judged in component based development due to the absence of the availability of the code of the component |
| V W X | Lack of Technology expertise and poor work knowledge and skills of assembler leads to Poor component evaluation and integration |
| Y Z A | Missing compatibility between the different versions of the component based software |
| A B C D | Existence of Poor or no documentation feature for the new versions |
| E F G H | If the stability is not incorporated in the component based system |
| IJ K L | Doing change in one component will make a heavy impact on the other component |
| M N O P | Unavailability of the competent staff |
| Q R S T | Unavailability of the internal structure of the component makes the testing process Tough and unreliable |
| A F | End user training sessions are not conducted |
| C F | Component based software that is developed cannot accommodate changes preferred by the use |

| B F | Occurrence of incompatibility between the component based product being developed and quality level that has been set during the initial phase of the software development cycle |
|---|---|
| E F | Complicated system manual results lack of understanding by the users |
| G F | Quality services after the COTS software installation at the user site are not given |
| K F | User is not facilitated with the upgraded copies of the component based software |
| L F | Updating or alteration of the component based system cannot be facilitated |
| M F | Lack of tracing of alternate component in case of failure [8] |
| T F | Planning the maintenance is difficult as the components have asynchronous cycle [8] |

## Determining the Total Risk Value of the Software

The Total Risk Value of the component based Software is equal to the sum of the Final Risk Value of each risk encountered in every phases of the software development
TRVS=∑FRVR of every risk encountered in each phase of the development
TRVS = 220.15+112.15+53.04+2`2.74
TRVS=408.08

## Determining the Risk Factor

Risk Factor (RF) = TRVS/ ITRVS
ITRVS is the Ideal Total Risk Value of the Software
ITRVS can be calculated when all the Risk encountered attain a rank of "3.5"by the Users and Risk Manager.

TRVS = 408.08

ITRVS = ITRVS of the users + ITRVS of the Risk manager

ITRVS=700+840
ITRVS=1540
RF = TRVS/ITRVS
RF = 408.08/1540

RF = 0.264

## Determining the Risk Severity from RF

If $0.0 \leq RF \leq 0.25$
Software Risk is "Negligible"
If $0.26 \leq RF \leq 0.50$
Software Risk is "Low"

If $0.51 \leq RF \leq 0.75$
Software Risk is "Moderate"
If $0.76 \leq RF \leq 1.00$
Software Risk is "High
Through this survey Risk factor come out to be in the second risk rating that reflects risk as low. Therefore software tends to be less risky

## 6. Conclusion

Our study explored the occurrences of several risks in COTS - based projects and in the RUP phases .We have discussed that how RUP entails COTS-based projects activities and the concurrency between the components based life cycle activities and the RUP phases. Risk attack on different component development phases vary with respect to the different nature of projects we have highlighted the several risks that cut across a component-based development cycle. Discussion like various phases of RUP associated with the several risks is also been incorporated. This research focused on the importance of Risk Identification in component based development. This identification has made the risks more visible at each component development stage making it possible to carry out activities that can minimize their effects. Detailed discussions on the issues like risk characteristics of the component based development have also been addressed. A survey of component based software is done based on the proposed risk characteristics in RUP Life cycle to evaluate its severity in terms of risk. We acknowledge that a lot has to be done in CBD. Future works involves validating and the impact of the risk reduction activities on the corresponding risks .Study regarding finding out the quantitative model for the risk analysis so that the risk can be calculated in no time.

## 7. Future Work

Furthermore, the validations and verification of components needs to be addressed .Special attention must be given towards the standardization of domain-specific components on the interface level that will lead to the development of the application components purchased from different vendors. CBSE is facing many challenges today. Questions like "if system attributes derivable from the component attributes" is still a subject of research. Queries related with the trustworthiness of the components are still unresolved. Effects of degrees of trustworthiness on system attribute unknown. Process models being used in component based development are still incomplete .Maintainability of the component based systems is still troublesome. Solutions of the updating of components dynamically is still the subject of research .CBSD is still facing the challenges of providing variety of tool support like test tools, configuration tools, evaluation tools etc .future work includes developing and evaluating certain tools for automating integration tests, that could be integrated in the protector's development process. We acknowledge that a lot more needs to be done in the area of CBD. Future work involves establishing a set of mitigation strategies for the risks identified during various component-based development activities to take advantage of COTS technology. Lot has to be done towards how modular risk assessment can applies to different risk modeling techniques .UML techniques needs to be extended in order to incorporate various component development phases

## References

[1] Abdullah,Tahir& Mateen, Ahmed& Raza, Ahsan& Mustafa,Tasleem, (2010) "Risk Analysis of Various Phases of Software Development Models", European Journal of Scientific Research ISSN 1450-216X Vol.40 No.3 , pp.369-376

[2] Johar,Kaur,amandeep & goel,shivani, (2011) "Cots Components Usage Risks In Component Based Software Development", International Journal of Information Technology and Knowledge Management, Volume 4, No. 2, pp. 573-575

[3] Rashid,Awais & Kotonya,Gerald ,(2011)"Risk Management in Component-based Development:A Separation of Concerns Perspective"

[4] Brændelanda,Gyrd&Refsdala,Atle&Stølen,Ketil, (2011), "A denotational model for component-based risk analysis", ISSN 0806-3036

[5]Brændelanda,Gyrd&Stølen,Ketil(2012),"Using Model-Driven Risk Analysis in Component-Based Development",IGI Global

[6] G. McGraw, "Software Security,",*IEEE* Security & Privacy, vol. 2, no.2, 2004, pp. 80–83.

[7] Vitharana,Padmal,(2003), "Risks and challenges of the component based software development", COMMUNICATIONS OF THE ACM, VOL. 46, NO. 8

[8] LI,Jingyue & CONRADI,Reidar & Petter,Odd& N. SLYNGSTAD & TORCHIANO,Marco& MORISIO, Maurizio & BUNSE,Christian, (2007)," A State-of-the-Practice Survey on Risk Management in Development with Off-The-Shelf Software Components"

[9] Mahmood,Sajjad & Lai,Richard, (2006)" Analyzing Component Based System Specification", AWRE Adelaide, Australia

[10]Goertzel,Mercedes,Karen & Winograd,Theodore, (2011), "Safety and Security Considerations for Component-Based Engineering of Software-Intensive Systems"

[11] Grady Booch, Ivar Jacobson, and James Rumbaugh, Unified Modeling Language 1.3, White paper,

Rational Software Corp., 1998.

[12]Sharma,Vidushi&Baliyan,Prachi,(2011),"Maintainability Analysis Of Component Based System",IJSEIA Journals,vol5_no3

[13]Selvi,R.Thirumalai&Balasubramanian,N.V&T.Manohar,George, (2008)," Framework and Architectural Style Metrics for Component Based Software Engineering", Proceedings of the International MultiConference of Engineers and Computer Scientists, Vol IIMECS 2008, 19-21 March, 2008, Hong Kong

[14] Johar,Kaur,amandeep & goel,shivani, (2011) "Designing of RIMCOTS model for Risk identification and mitigation for COTS-based Software Development", International Journal of computer system engineering, Vol 02, Issue 02, pp. 573-575, ISSN: 2230-8563; e-ISSN-2230-8571

[15] Kotonya, G& Rashid, A, (2001)," A strategy for managing risk in component-based software development ",ISBN: 0-7695-1236-4 2001

[16] Sagredo,Victor& Becerra, Carlos& Valdes,Gonzalo, (2010),"Empirical Validation of Component-based Software Systems Generation and Evaluation Approaches", CLEI ELECTRONIC JOURNAL, VOLUME 13, NUMBER 1, PAPER 6

[17] Sharma,Arun & Kumar,rajesh & Grover,P.S, (2007)," A Critical Survey of Reusability Aspects for

Component-Based Systems",WASET JOURNAL,v33-8